\RequirePackage{fix-cm}
\documentclass[twocolumn]{svjour3}          
\smartqed  
\usepackage{graphicx}
\usepackage{color,soul}
\usepackage[numbers]{natbib}
\begin{document}

\title{Micro-mechanical Failure Analysis of Wet Granular Matter}
\author{Konstantin Melnikov$^1$ \and
        Falk K. Wittel$^1$ \and 
        Hans J. Herrmann$^1$.
}

\institute{Konstantin Melnikov\at
 $^{1}$Computational Physics for Engineering Materials,\\ 
 ETH Zurich, Stefano-Franscini-Platz 3, CH-8093 Zurich, Switzerland \\
 \email{konstantin.melnikov@ifb.baug.ethz.ch} }
\date{Received: date / Accepted: date}
\maketitle

\begin{abstract}
We employ a novel fluid-particle model to study the shearing behavior of 
granular soils under different saturation levels, ranging from the dry material 
via the capillary bridge regime to higher saturation levels with percolating 
clusters. The full complexity of possible liquid morphologies 
\cite{Scheel2008nature} is taken into account, implying the formation of 
isolated arbitrary-sized liquid clusters with individual Laplace pressures that 
evolve by liquid exchange via films on the grain surface \cite{Melnikov2015}. 
Liquid clusters can grow in size, shrink, merge and split, depending on local 
conditions, changes of accessible liquid and the pore space morphology 
determined by the granular phase. This phase is represented by a discrete 
particle model based on Contact Dynamics \cite{Brendel2005}, where capillary 
forces exerted from a liquid phase add to the motion of spherical particles. We 
study the macroscopic response of the system due to an external compression 
force at various liquid contents with the help of triaxial shear tests. 
Additionally, the change in liquid cluster distributions during the compression 
due to the deformation of the pore space is evaluated close to the critical 
load.

\end{abstract}
\section{Introduction}\label{intro}
The presence of liquid is known to have a significant impact on the mechanical 
properties of granular materials \cite{Herminghaus2005,Mitarai2006}. An 
illustrative example we know from our childhood are sand castles: while it is 
nearly impossible to build with dry sand, one can easily accomplish this task 
after mixing some water to it. However, if we keep on adding more water, from a 
certain saturation on, the structures collapse. Cohesion by capillary forces is 
the reason for the change in material behavior. After reaching a maximum at 
lower saturation \cite{Scheel2008nature}, cohesive forces slowly decrease until 
the gravitational load gets too large. The nature of this phenomenon can be 
studied by tracking the liquid distribution on the micro-scale. In the last 
years, advanced experimental tools like micro-tomography created this 
possibility and Scheel \textit{et al.} found a rich variety of liquid cluster 
morphologies \cite{ScheelPHD,Scheel2008nature} beyond the well-studied liquid 
bridge regime \cite{Mani2012,Mani2012b,Richefeu2006,Scholtes2009}. The number 
and the size of observed liquid clusters proved to strongly depend on the 
saturation level. Inspired by these findings, we developed a model that 
explicitly takes into account all possible liquid morphologies on the pore scale 
inside a random sphere packing \cite{Melnikov2015}. Better understanding of the 
impact of higher liquid content is crucial for solving a number of open problems 
including e.g. rainfall-induced slope failures or fluid motion in sheared 
granular systems. 

Wet granulates at low saturation, i.e. being in the so-called pendular state, 
are well studied 
\cite{Groeger2003,Mani2012,Richefeu2006,Scholtes2009a,Scholtes2009}. With 
regard to cohesion the fully saturated state resembles the dry one since all cohesive 
forces vanish. However 
intermediate states, as well as the evolution of liquid structures under 
imbibition and drainage are rarely addressed due to their complexity. Starting 
from our grain-scale model for arbitrary saturation \cite{Melnikov2015}, 
 we now allow particles to move and introduce a coupling between fluid and 
particles. Our intention is to investigate the effect of higher saturation on 
the macroscopic behavior of the material and vice versa the effect of 
macroscopic deformations on the morphogenesis of the liquid body. To capture the 
effect of hydrostatic stress, triaxial shear tests are simulated at different 
initial saturation states \cite{Belheine2008,Wang2001}.

First we give a short description of the model with respect to the used discrete 
particle method (DPM) of contact dynamics (CD), before we summarize the most 
important features of the implemented liquid structure and evolution model. 
Special attention is given to the coupling between fluid and particles. The 
numerical procedure is described in detail and the triaxial shear cell used for 
measuring macroscopic behavior is introduced. We discuss the macroscopic 
response of the granular material to external compression for different 
saturation levels and the respective observations of the liquid body before we 
summarize main results and draw conclusions in Sec.~\ref{sec:Conclusions}.
\section{Fluid-Particle Model for Arbitrary Liquid Saturation}\label{sec:1}
We focus on wet granular material at various saturation levels under triaxial 
compression. Consequently one has to consider a solid, liquid and gaseous phase 
in the model. We limit ourselves to granular material consisting of uni-sized 
spherical particles and simulate time evolution of the dense packing under 
external load by Contact Dynamics. For the liquid phase, we develop a 
grain-scale model, capable of resolving the liquid body by combinations of 
capillary bridges, menisci and fully saturated pores. This way we resolve local 
liquid clusters from the dry to the fully saturated state. However gas can be 
trapped inside of pores, that is considered as incompressible.
\subsection{Solid Phase Model}\label{Sec:DEM}
The solid phase is modeled using a DPM based on Contact Dynamics (CD). 
Originally proposed by Moreau~\cite{Moreau1994,Brendel2005}, this method is 
particularly suited representing rigid frictional particle systems with high 
packing fractions but negligible particle elasticity. Hence Newton`s equations 
of motion are integrated implicitly, considering a volume exclusion constraint 
to avoid overlap of particles. In such systems, the rearrangements of rigid, 
frictional particles exclusively determines the system's behavior.

\begin{figure}[htb]
\centering
\includegraphics[]{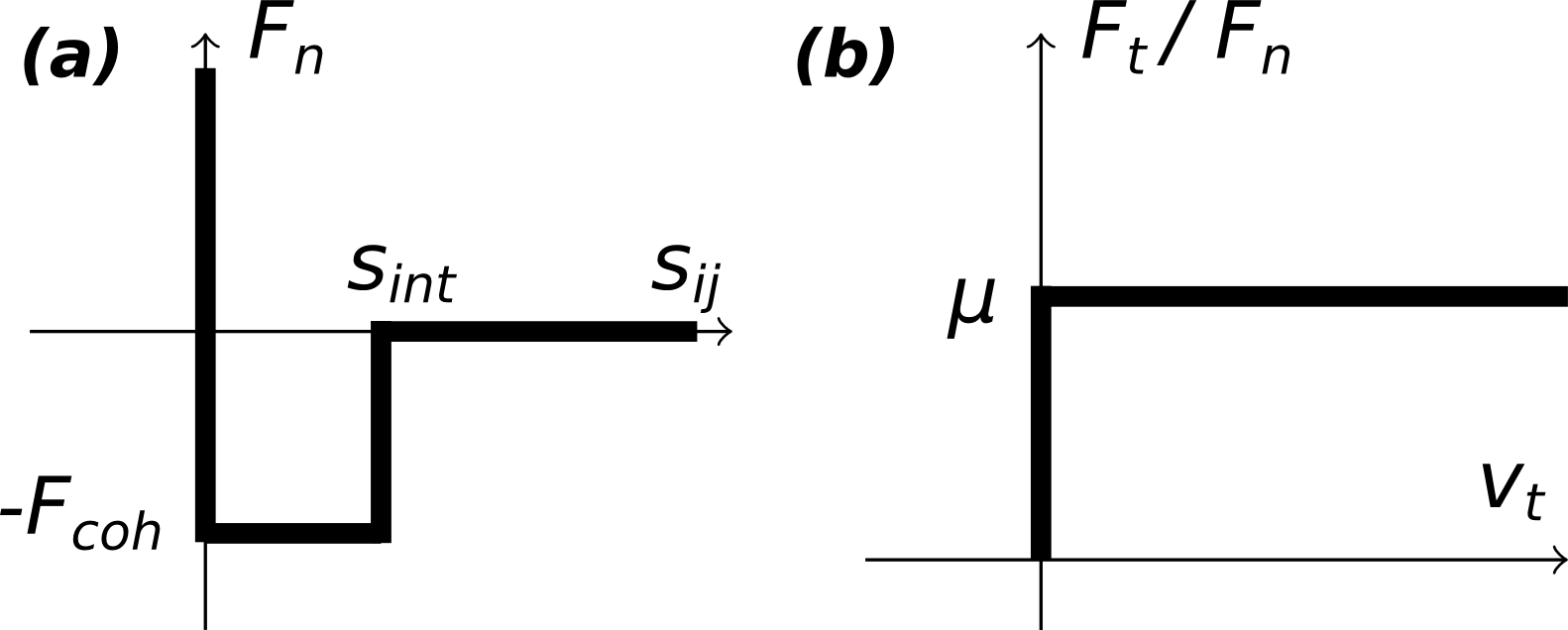}
\caption{Contact laws in Contact Dynamics. (a) Repulsive normal force $F_n$ with 
volume exclusion constraint for $s_{ij}=0$ and attractive cohesive force 
$F_{coh}$ with interaction distance $s_{int}$. (b) Sliding of particles is 
suppressed until the tangential force $F_t$ is smaller than the Coulomb friction 
threshold $\mu F_n$.\label{Fig:contactLaw}}
\end{figure}

Each particle experiences a force $\vec{F_p}$ that is composed of the contact 
forces $\vec{F_c}$ with all contacting particles and the wall, a gravitational 
force $\vec{F_{ext}}$, as well as forces from the liquid, that are described 
along with the considered liquid structures in Sec.~\ref{sec:liquidbody}. The 
essential difference between CD and soft particle methods is the strict volume 
exclusion constraint, where the normal force component $F_n$ can take arbitrary 
large positive values if the distance or gap $s_{ij}=0$ between surfaces of 
particle $i$ and $j$, to prevent interpenetration of surfaces (see 
Fig.\ref{Fig:contactLaw}). Hence, in an iterative process the position of the 
particle must be found for which the normal force component $\vec{F_n}$ of each 
active contact is minimal for all active contacts such that there is no overlap 
of particles at the next time step. When $s_{ij}>0$, the corresponding $F_n$ 
vanishes, unless attractive cohesive forces $-F_{coh}$ act in the normal 
direction, e.g. in a range $0<s_{ij}<s_{int}$. Sliding between particles is 
prevented if the tangential component of the contact force $F_t$ does not exceed 
the threshold given by the Coulomb friction coefficient $\mu$: $0 \leq F_t \leq 
\mu F_n$ (see Fig.~\ref{Fig:contactLaw}(b)). Note that the static friction 
coefficient is set equal to the dynamic one. The tangential force at the contact 
point and the resulting moment around the particle center lead to particle 
rotation.

In a dense packing, many particles are interacting in a force contact network. 
The contact forces acting on each particle are influenced by adjacent contact 
forces in the neighborhood, requiring a global solution of contact forces. The 
ones for a particle are described by a system of equations which can be 
approximately solved by an iterative algorithm on all particles in a random 
order. Over several iterations, the global contact force network relaxes towards 
a state where only minor changes occur. The calculated forces are then used to 
integrate the equations of motion by an implicit Euler method. The larger 
computational cost for the iterative force calculation is often outbalanced in 
CD by the possibility to use significantly larger time steps compared to soft 
particle dynamics.
\subsection{Liquid Phase Model}
From micro tomography experiments it is known, that liquid builds highly complex 
structures inside the granular material above the liquid bridge regime 
\cite{Scheel2008nature}. In order to simulate this behavior, we use a general 
model for liquid saturation which is able of resolving all possible liquid 
structures observed in experiments. In this section we give a short overview of 
this model, while the detailed description can be found in 
Ref.~\cite{Melnikov2015}. 

\begin{figure*}[htb]
\centering
\includegraphics[width=14cm]{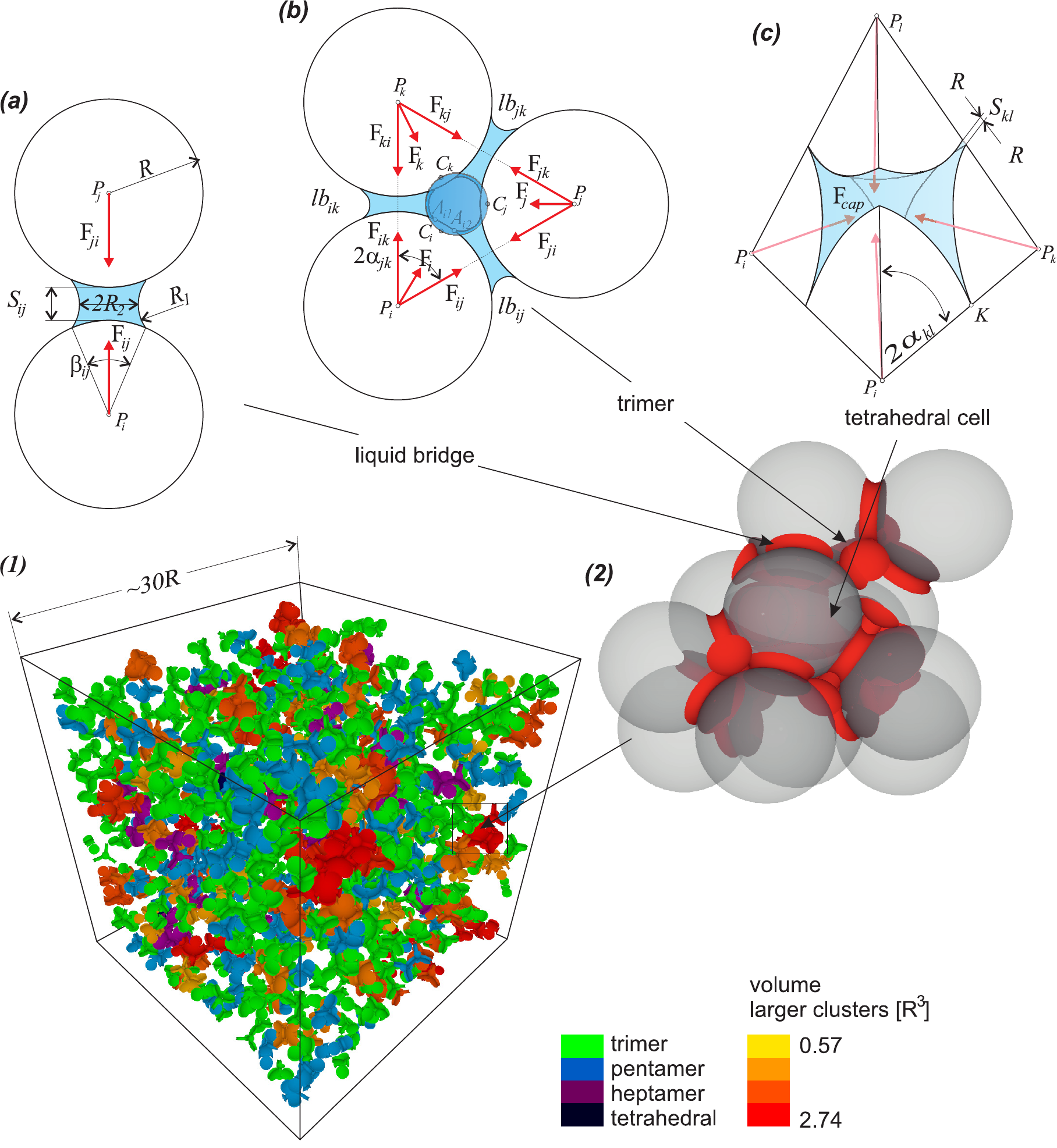}
\caption{\label{Fig:model} (Color online) Liquid clusters emerging in a sphere 
packing at liquid content $W_c=3\%$ $(1)$ and one exemplary cluster $(2)$, 
composed of three basic liquid structures: bridges $(a)$, trimers $(b)$, and 
tetrahedral cell $(c)$ with definitions given in the text. Note that in $(b)$, 
$C_i$ is located above the trimer plane. } 
\end{figure*}

Based on the exact geometrical positions of the particles, the pore-throat 
network is constructed via Delaunay triangulation. The void space in each 
tetrahedral cell of the triangulation (see Fig.~\ref{Fig:model}(c)) is called 
the pore body, while the cutting areas of the pore body with the respective 
faces of the cell form the four pore throats. Liquid clusters are represented as 
a combination of three types of elementary units: \textbf{liquid bridges}, 
\textbf{menisci} and entirely filled \textbf{pore bodies}. These elements can 
form higher geometrical configurations called \textbf{liquid clusters}. This 
evolution is determined by local instability criteria for imbibition and 
drainage.
\subsubsection{Basic Liquid Structures}\label{sec:liquidbody}
The elemental liquid structure involving two grains $i$ and $j$ of radius $R$, 
is a \textbf{liquid bridge} (Fig. \ref{Fig:model}(a)). With their centers 
$P_{i},P_{j}$ and $R$, the inter-particle separation distance $S_{ij}$ is given. 
In general, the pressure drop $\Delta P$ = $P_{liquid} - P_{gas}$ across the 
liquid-gas interface with surface tension $\gamma$ and curvature $C=1/R_1+1/R_2$ 
with the principal radii of curvature $R_{i,j}$, is described by the 
Young-Laplace equation \cite{deGennes2003}: $\Delta P=\gamma C$. We interpolate 
the solution of this equation for liquid bridges from tabulated values by 
Semprebon \textit{et al.}~\cite{Mani2015} obtained with a numerical energy 
minimization method (NEM) of the software Surface Evolver \cite{Brakke1996} for 
a constant contact angle $\theta=5^{\circ}$. With the liquid volume of the 
bridge $V_{ij}$, being a state variable, we interpolate $\Delta 
P_{ij}(R_i/R_j,V_{ij},S_{ij})$, as well as the filling angle 
$\beta_{ij}(R_i/R_j,V_{ij},S_{ij})$ from the tables. The capillary force of the 
liquid bridge $\vec{F}_{ij}$ acting on particles $i$ and $j$ in the direction 
given by the unit vector $\vec{e}_{ij}$ from $P_i$ to $P_j$ is calculated via 
the experimentally found expression by Willett \textit{et 
al.}~\cite{Willett2000}
\begin{equation}
 \vec{F}_{ij} = -\vec{F}_{ji} = \vec{e}_{ij}\cdot\frac{2\pi R \gamma cos 
\theta}{1+0.5 S_{ij} \sqrt{R/V_{ij}} + 2.5 S_{ij}^{2} R/V_{ij}}.
 \label{Fcap}
\end{equation}
Note that for contact ($S_{ij}=0$) the capillary force is independent of 
$V_{ij}$, resulting in a constant force $\left|F_{ij}\right|$, provided 
$\gamma=const$ and $\theta=const$.

\begin{figure}[htb]
\centering
\includegraphics[width=8cm]{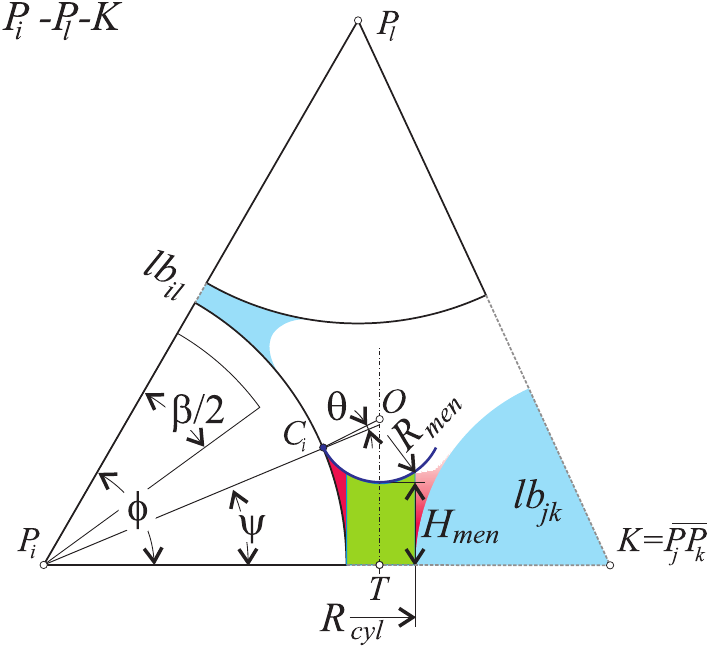}
\caption{\label{Fig:instab} Sectional view of tetrahedral cell through 
$P_i$-$P_{l}$-$K$ (Fig.~\ref{Fig:model}) showing the meniscus position in the 
pore throat. Point $O$ defines  the center of the spherical cap shaped meniscus 
of radius $R_{men}$. The green area below approximates the meniscus volume by a 
cylinder of radius $R_{cyl}$ and height $H_{men}$, while the red area points at 
the underestimated volume.}
\end{figure}

As soon as three particles $i,j,k$ with respective liquid bridges 
$lb_{ij},lb_{ik}$ and $lb_{jk}$ are involved, the remaining pore throat can be 
filled, on condition that enough liquid is stored in the bridges, resulting in 
trimer formation (Fig.~\ref{Fig:model}(b)). Based on ideas by 
Haines~\cite{Haines1927}, we model the resulting liquid body of equal Laplace 
pressure by adding to the three  bridges a fluid filled cylindrical pore throat 
that is located between the grains and limited by two \textbf{menisci} of 
spherical shape as liquid-gas interface (see section cut Fig.~\ref{Fig:instab}). 
The position of the meniscus center $O$ and its radius $R_{men}$ is calculated 
following the geometric construction proposed by Gladkikh \cite{GladkikhPHD} 
with contact angle $\theta$. $O$ and $R_{men}$ determine the fluid volume in the 
cylindrical pore throat $V_{cyl}$ of radius $R_{cyl}$ and height $H_{mem}$. 
$C_i$ denotes the contact point between meniscus and particle $i$ (see 
Figs.~\ref{Fig:model}(b),\ref{Fig:instab}). The negative pressure drop $\Delta 
P$ across the meniscus is calculated using the Young-Laplace equation with the 
constant curvature $C=1/R_{men}$. A capillary force $\vec{F}_{i,j,k}$ from the 
meniscus acts on particles $i,j,k$ additionally to the forces from the liquid 
bridges. For simplicity, it acts in the direction given by the points 
$\overline{P_iT}$ (see Fig.~\ref{Fig:instab}) with the unit vector 
$\vec{e}_{iT}$ and is calculated as $\vec{F}_{i,j,k}=\vec{e}_{iT} \cdot \Delta P 
A_{eff}$ with the effective wetted area $A_{eff}$ defined for example for 
particle $i$ by a triangle through the points $C_i$-$A_{i1}$-$A_{i2}$ (in 
Fig.~\ref{Fig:model}). 

With four particles, a tetrahedral cell can be formed, with the void space in 
the center being filled by a \textbf{pore body}. This pore body can be either 
connected to another filled pore body or it can be bounded by a meniscus through 
each of the four existing pore throats. In the latter case, the meniscus defines 
the Laplace pressure of the liquid phase inside the pore. The Laplace pressure 
drop $\Delta P$ leads to a capillary force component due to the wetted particle 
surface area inside the pore body. The resulting capillary force $\vec{F}_{cap}$ 
pulls each of the four particles towards the center of the pore as shown in 
Fig.~\ref{Fig:model}(c). The respective wetted particle surface area is the 
surface of each particle inside the triangulation cell. If a liquid bridge 
between the corresponding particles exists, their wetted area is reduced in the 
calculation by the wetted area of the liquid bridge with opening angle 
$\beta_{ij}$. Note that in this case we do not correct the direction of the 
force and for simplicity keep it directed from the particle to the cell center. 
When particles are entirely immersed in liquid, forces from the pore body 
vanish, since the pressure inside the cluster is constant. When a tetrahedral 
cell contains four menisci, an incompressible gas bubble can be trapped.

The elementary units liquid bridge, meniscus, and filled pore body are the 
building blocks of local \textbf{liquid clusters} which can evolve inside the 
granular material (see Fig.~\ref{Fig:model}(1-2)). Following the description of 
Scheel \textit{et al.}~\cite{Scheel2008nature} for clusters without filled pore 
bodies, we call the smallest possible cluster trimer (three contacts, see 
Fig.~\ref{Fig:model}b), the next one is a pentamer, heptamer and if more than 7 
contacts are involved we simply call it cluster. The smallest cluster with a 
filled pore body is a tetrahedral cluster consisting of 6 associated liquid 
bridges and 4 menisci. We assume that pressure inside a each liquid cluster is 
constant,  implying that all menisci have the same curvature. Pressure is 
calculated individually for each liquid cluster based on its volume 
\cite{Melnikov2015}, which is the control variable of the model. With $N_{imb}$ 
imbibed pore bodies and $N_{men}$ menisci in a cluster
\begin{equation}
V_{c}(R_{men}) = \sum_{i=0}^{N_{imb}} V_{pore, i} + \sum_{j=0}^{N_{men}} V_{men, 
j}(R_{men}), 
\end{equation}
where $V_{pore, i}$ denotes the volume of the imbibed pore body $i$ and $V_{men, 
j}$ the volume of the meniscus $j$ including its associated liquid bridges. The 
volume of the filled pore is calculated by subtracting the partial volumes of 
the four particles contained within the tetrahedral cell from the volume of this 
cell. The second component in the term $V_{men}(R_{men})$ accounts for the 
volumes of the liquid bridges $V^{lb}_{i}$ associated with the meniscus $i$:
\begin{equation}
V_{men}(R_{men}) = V_{cyl}(R_{men}) + (0.5+\varepsilon)  \sum_{i=0}^{m} 
V_{i}^{lb}(R_{men}).
\label{eq:menVol}
\end{equation}
The geometrical correction parameter $\varepsilon$ describes the volume excess 
of a real meniscus with the connected liquid bridges, compared to our 
approximation in which $V_{cyl}$ underestimates the volume bounded by the 
meniscus (see Fig.~\ref{Fig:instab} red area and Ref. \cite{Melnikov2015}). Only 
half of each bridge volume is considered in the above formula since the other 
half is located in the opposite triangulation cell.
\subsubsection{Evolution of Liquid Structures} \label{sec:stabilitCriteria}
Liquid structures can change due to relative particle movements or volume 
changes, for example through condensation or evaporation at gas-liquid 
interfaces or by local liquid sources or sinks. To take recent experimental 
observations into account \cite{Lukyanov2012,Scheel2008nature}, we include 
transport via the liquid film on the particle surface. This transport
occurs at low and intermediate saturations, where liquid 
exists in form of spatialy disconnected clusters.
For stationary flow in 
the film, the volume flux $\dot{V_{i}}$ into structure $i$ is proportional to 
the local Laplace pressure differences between liquid structures sharing at 
least one grain. If $N_i$ structures can be connected to the same grain, one 
obtains
\begin{equation}
\label{eq:liqu_trans}
\dot{V_{i}} = \frac{R}{\gamma}\cdot\sum_{j=0}^{N_i} \omega_{ij} (P_{j} - P_{i}),
\end{equation}
with the dimensionless conductance coefficient $\omega_{ij}$ that defines the 
equilibration time scale. In general $\omega_{ij}$ should depend on details like 
distance between structures, the number of structures connected to the grain and 
others \cite{Mani2015,Mani2014}. However for simplicity we set 
$\omega_{ij}=0.01$ for all presented simulations (see 
Tab.~\ref{Tab:parameters}). Particle movement, as well as flow from or into 
liquid structures can trigger local instabilities that propagate liquid 
interfaces inside the porous particle packing. This is consequently a 
discontinuous process with instantaneous jumps between stable configurations. 
Micro mechanically, these interface jumps are associated with drainage or 
imbibition of pore bodies or throats, or bridge rupture. We identified a set of 
seven geometrical criteria for instability, similar to Motealleh \textit{et 
al.}~\cite{Motealleh2013} and Gladkikh~\cite{GladkikhPHD}, with criteria 
\textbf{i1-i4} for imbibition and \textbf{d1-d3} for drainage:

\textbf{Criterion i1:} If two liquid bridges with filling angles $\beta_{ij,ik}$ 
touch each other, they can form a new trimer by filling the pore throat if 
sufficient liquid for a stable meniscus according to the drainage criterion 
$\textbf{d1}$ is available. With the opening angle $\alpha_{jk}$ (see 
Fig.~\ref{Fig:model}(b)) it reads $0.5(\beta_{ij}+\beta_{ik})>2\alpha_{jk}$.
   
\textbf{Criterion i2:} If a meniscus in a tetrahedral cell touches a single 
liquid bridge, that was up to now not part of the liquid body, the pore body is 
filled. Hence the respective meniscus with the filling angle of the meniscus 
$\psi$ and the face-edge angle $\phi$ shown in Fig.~\ref{Fig:instab} becomes 
unstable when $\psi+\beta/2>\phi$.
   
\textbf{Criterion i3:} If the centers of two menisci $O$ inside one pore body 
touch, they build a single spherical interface and can fill the pore body (see 
Fig.~\ref{Fig:instab}), however with all four menisci formed, a gas bubble gets 
trapped.
   
\textbf{Criterion i4:} If a meniscus touches the opposite particle, the pore 
body is imbibed.
   
\textbf{Criterion d1:} Pore throats will drain if they reach a minimum size, 
expressed by the critical height of the menisci $H_{men}^{min} \geq \kappa R$ 
with the arbitrary drainage parameter $\kappa$. Best agreement with experimental 
data was found for $\kappa=0.15$ \cite{Melnikov2015}.

\textbf{Criterion d2:} Pore bodies become unstable when the center of the 
meniscus of a neighboring cell touches the respective pore throat plane of an 
entirely saturated cell. Instantaneously the liquid interface jumps to a new 
stable position, leaving behind a drained pore body.

\textbf{Criterion d3:} Liquid bridges between particles $i$ and $j$ can rupture 
when the empirical expression derived by Willett \textit{et 
al.}~\cite{Willett2000} for the rupture distance $S_{c} \simeq 
(1+\frac{1}{2}\theta)(\sqrt[3]{V_{ij}}+\sqrt[3]{V_{ij}^2}/10)$ in units of the 
particle radius $R$ to the bridge volume $V_{ij}$ is reached (see 
Fig.~\ref{Fig:model}(a)). $S_c$ in combination with a zero distance for the 
formation of liquid bridges results in the hysteretic behavior of wet granular 
material. $V_{ij}$ is sucked back to the surface of the holding particles and 
redistributed equally to their other contacts \cite{Mani2012b}.
\subsection{Numerical Procedure}
\begin{figure*}[htb]
\centering
\includegraphics[width=12cm]{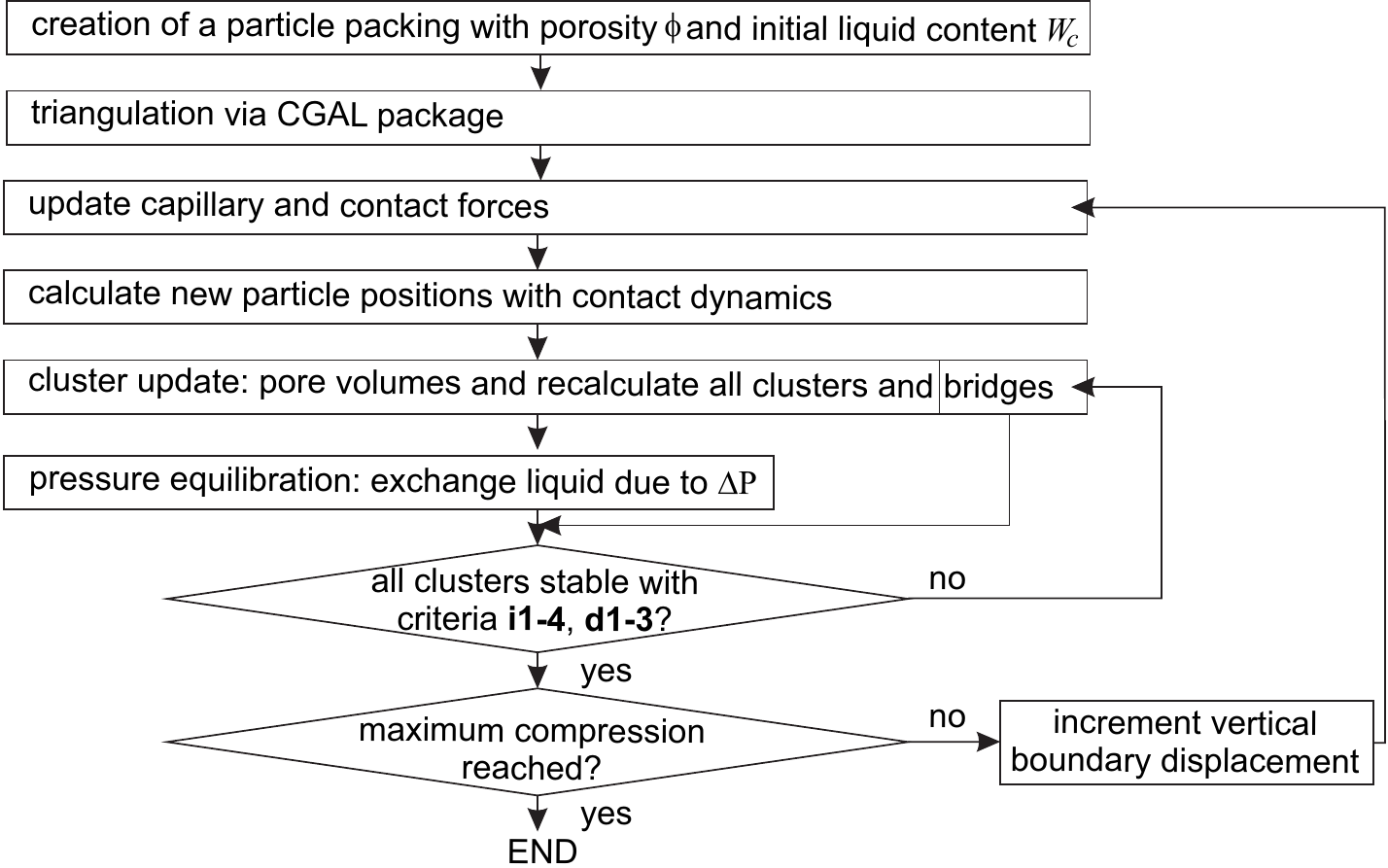}
\caption{\label{Fig:flowdia} Simulation scheme.} 
\end{figure*} 
The different steps of the numerical procedure are schematized in 
Fig.~\ref{Fig:flowdia}, while a detailed description of the system construction 
and fluid calculation was previously published in Melnikov \textit{et 
al.}~\cite{Melnikov2015}. The initial packing, that is dynamically created via 
the DPM model (see Sec.~\ref{Sec:DEM}) already allows for the introduction of an 
initial liquid content by assigning a liquid film to each particle, small enough 
to stay in the pendular regime \cite{Herminghaus2005}. When particles contact, 
liquid bridges are created that can rupture under criterion \textbf{d3}.We define
the liquid content $W_c$ as total liquid volume $V_{liquid}$ normalized by the 
void volume $V_{void}$: $W_{c} = V_{liquid} / V_{void}$, where the void volume
$V_{void}$ is calculated as the difference between the sample volume $V_{sample}$ and 
the total volume of the grains $V_{grains}$. The 
porosity $\phi$ is defined as $\phi=1-\phi_s$, where $\phi_s$ is the solid 
fraction given by the total volume of $N_p$ particles via $\phi_s=4N_p\pi 
R^3/(3V_{sample})$. Before entering the simulation, the pore space is 
triangulated into tetrahedral cells using a Delaunay triangulation on particle 
centers \cite{GGAL_Periodic3D}. To increase $W_c$, liquid is condensed into 
already existing bridges, leading to formation of local clusters due to 
occurring instabilities. To avoid boundary effects with the wall, we suppress 
the formation of liquid structures at a distance $d \approx 2R$ from all walls. 
After addition of liquid to existing structures, we recalculate them such that 
the Laplace pressure corresponds to the changed volume. Before checking for 
instabilities, the liquid transport due to Laplace pressure gradients is 
calculated for the entire time increment, leading to equilibration of pressure 
differences. Then the instability criteria for imbibition \textbf{i1-i4} and 
drainage \textbf{d1-d3} are checked in ascending order. Identified instabilities 
are eliminated through drainage or imbibition as described in 
Sec.~\ref{sec:stabilitCriteria} and a cluster update is made before the time is 
incremented by $\Delta t$. During the new time step, we first update particle 
positions through the DPM algorithm, and consecutively calculate the respective 
liquid body configuration as described above. To ensure conservation
of the total liquid volume in the specimen, all 
liquid clusters are subsequently recalculated considering the changed volume of 
the pore bodies. Note, that in this paper we stick to small strains up to 4$\%$ 
to be able to retain the initial pore-throat network.
\subsection{Triaxial Shear Test Model}\label{sec:Setup}
\begin{figure}[htb]
\centering
\includegraphics[width=8cm]{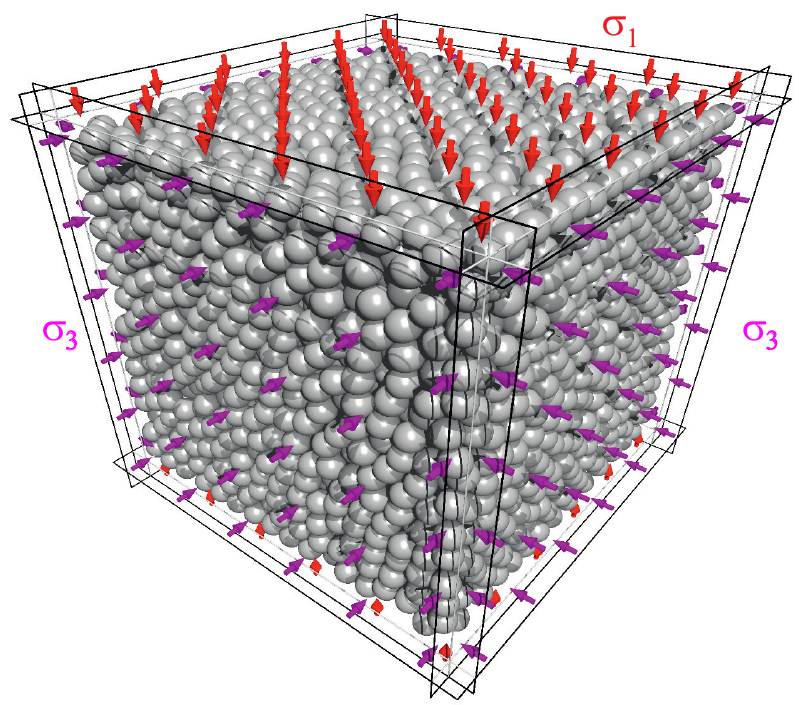}
\caption{Setup of the triaxial shear test with confining walls. 
\label{Fig:triaxialSetup}}
\end{figure}

To study the shear strength of the wet granulate, we simulate triaxial shear 
tests that are a common tool in geotechnique for obtaining macroscopic systems 
responses \cite{Belheine2008,Scholtes2009}. The cubic sample contains $N_p$ 
randomly distributed particles of mass density $\rho_{p}$, radius $R$ and hence 
particle mass $m_p$. To randomize the particles, the solid fraction is kept low, 
and random velocities are assigned to the particles, before the sample is 
confined by a hydrostatic pressure on each cube walls of mass 2000$m_p$. To 
obtain the desired solid fraction $\phi_s=0.6$, friction is omitted, resulting 
in initial system dimensions of $L_{x,0}=L_{y,0}=L_{z,0} \approx 32 R$ (see 
Fig.~\ref{Fig:triaxialSetup}). Note that in CD overlaps are avoided. Hence no 
elastic energy is stored in overlaps, opposite to soft particle dynamics. The 
confining forces on each wall $F_{conf}$ are calculated with the respective 
stress $\sigma_{1,3}$ and the true area of the wall $A_{wall}$ involved. 
$\sigma_{1}$ and $\sigma_{3}$ denote the confining pressures at the upper and 
side walls, respectively (see Fig.~\ref{Fig:triaxialSetup}). For the compression 
with $\sigma_1 \neq \sigma_3$ friction is switched on, and the mass of the 
bottom wall is set to infinity. The hydrostatic stress $\sigma_3$ is applied on 
all movable walls before the upper wall of the cubic sample is lowered at a 
constant strain rate $\dot{\nu}=v_{w}/L_{z}$ with the vertical displacement 
rate $v_{w}$. The strain rate $\dot{\nu}$ is chosen such that the assumption 
of quasi-static compression is satisfied, meaning that the response of the 
system for even lower strain rates would be the same. The unit of time $T$ is 
defined as $T = 1/\dot{\nu}$ meaning that for a simulation with
$\dot{\nu}=0.0001$ $[1/T]$ the upper wall moves by a distance of $0.0001L_z$ 
during a time unit. 

We calculate the differential stress $q=\sigma_{1}-\sigma_{3}$ by measuring the 
normal stresses on the upper and the side walls from the reaction forces of the 
particle contacts. The axial strain is calculated via $\varepsilon_{A}=\mid 
L_z-L_{z,0}\mid / L_{z,0}$. Note, that we use the absolute value in the above 
definition which implies that the axial strain during compression is positive. 
The volumetric strain $\varepsilon_V$ is calculated as $\varepsilon_V=\Delta V / 
V_0$ where $\Delta V$ is the volume change of the sample with respect to the 
initial volume $V_0$.
\begin{table}[htb]
\caption{\label{Tab:parameters} Parameters of the solid and liquid phase model.}
\begin{tabular}{lll}
 \multicolumn{3}{l}{\textbf{Solid phase:}} \\
    $N_p$ & 5000 & number of particles\\
    $R$ & 1 & particle radius\\
    $\rho_{p}$ & 1 & particle mass density\\
    $\phi_s$ & 0.60 & solid fraction\\
    $\Delta t$ & 0.001 - 0.01 & time step\\
    $\theta$& $5^{\circ}$ & contact angle \\   
    $\mu$ & 0.3 & friction coefficient \\   
 \multicolumn{3}{l}{\textbf{Liquid phase:}} \\
    $\omega$ & 0.01 & conductance coefficient\\
    $\varepsilon$& 0.07 & geometrical correction parameter\\
    $\kappa$ & 0.15 & drainage parameter for meniscus\\     
   $\gamma$& 1 & surface tension\\ 
\end{tabular}{}
\end{table}
\section{Results of Wet Triaxial Shear Tests}\label{sec:ShearTestMacroscopic}
To quantify the effect of liquid on the behavior of granular material under 
load, triaxial compression tests were simulated with  increasing initial liquid 
content $W_c$ from dry (0\%) up to 40\% and at three distinct values of 
confining pressure is $\sigma_3=20,40,60$ with the unit $u_p=m_p/(T^2R)$. Since 
no dimensional quantities are used in Contact Dynamics, a direct comparison with 
experimental data can be done by considering the ratio of inertial to confining 
forces proposed by Rognon \textit{et al.} \cite{Rognon2006}: $I=\dot{\nu} R 
\sqrt{\rho_p/P}$ where $P=\sigma_3$ is the confining pressure. In our 
simulations this number is equal to $I_{20}\approx 2.2 \cdot 10^{-5}$, 
$I_{40}\approx 1.6 \cdot 10^{-5}$ and $I_{60}\approx 1.3 \cdot 10^{-5}$. Note, 
that these inertial numbers correspond to the quasi-static regime since our 
simulations show no changes if the strain rate $\dot{\nu}$ is further 
reduced.
The resolution of local liquid structures gives the opportunity to study the 
evolution of liquid clusters during compression. We are interested in 
micro-mechanical changes in terms of morphology distributions of the liquid body 
or morphogenesis of individual clusters due to deformation. Finally we quantify 
the dependence of macroscopic dilatation and limit surfaces upon the liquid 
content.
\subsection{Liquid Cluster Evolution During Compression} 
\label{sec:ClusterEvolution}
\begin{figure}[htb]
\centering
\includegraphics[width=0.5\textwidth]{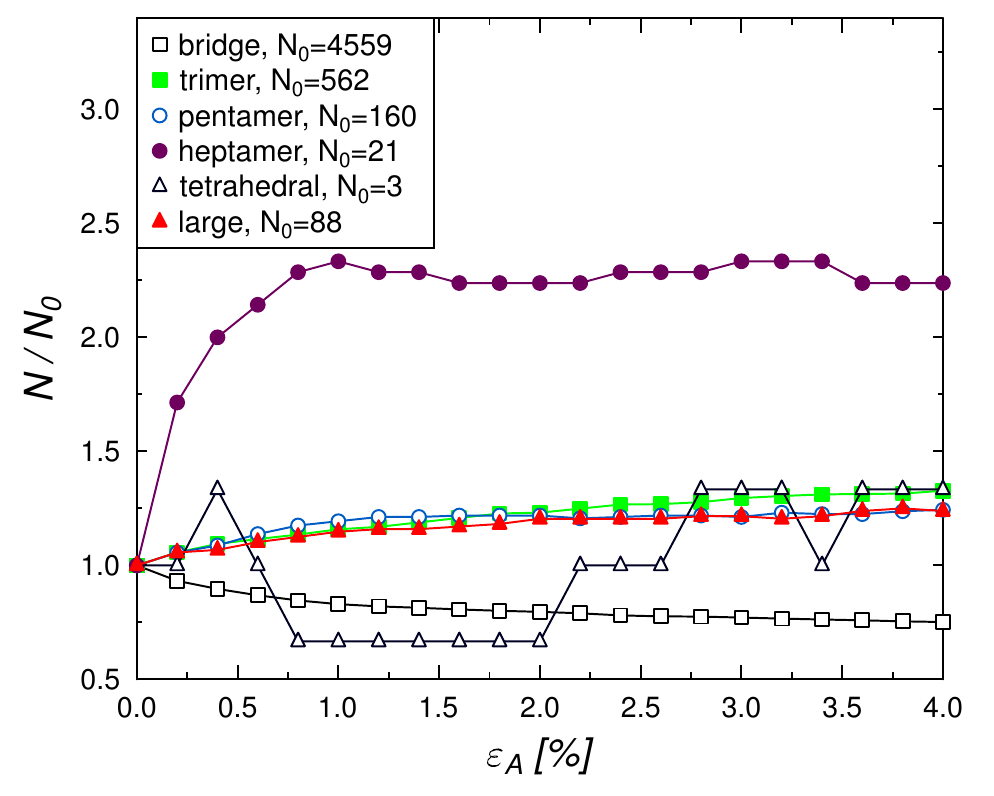}
\caption{Evolution of the liquid cluster morphologies $N$ scaled by $N_{0}$ at 
$t=0$ as a function of the axial strain $\varepsilon_A$. Liquid content 
$W_c=6\%$, normal stress on the vertical walls $\sigma_3=40 u_p$. 
\label{Fig:clusterStatistics}}
\end{figure}

In particulate systems, shear typically results in dilatation, that can strongly 
effect the stability of the liquid body. For a sample with liquid content 
$W_c=6\%$ and confining pressure of $\sigma_3=40 u_p$ we analyze the 
consequences of the compression on the distribution of liquid cluster 
morphologies from the single bridge up to large clusters. In 
Fig.~\ref{Fig:clusterStatistics} the relative change of the number of liquid 
cluster morphologies $N/N_0$ with respect to the stress-free state at $t=0$ 
(Fig.~\ref{Fig:model}(1)) is given as function of the axial strain up to 
$\varepsilon_A=4\%$. The compression in one axial direction as considered in 
this study leads to the dilatation of the sample in directions perpendicular to 
this axis. A drop in solid fraction and increasing sample volume are the 
consequences. Since dilatation means increasing inter-particle distances, it is 
not surprising that isolated liquid bridges become less common with increasing 
strain due to bridge rupture. The liquid is used for the formation of new 
trimers and of other cluster morphologies like pentamers and large clusters. 
Because of the limited sample size with $N=5000$, the total number of larger 
cluster morphologies is consequently small, resulting in strong fluctuations for 
tetrahedral clusters and heptamers. Nevertheless, one can conclude that the pore 
space deformation triggers new instabilities that lead to trimer formation and 
consolidation of existing clusters into larger structures. This process is 
fueled by the liquid volume from existing liquid bridges as their number drops 
and from other existing structures by coalescence. 

\begin{figure}[htb]
\centering
\includegraphics[width=0.48\textwidth]{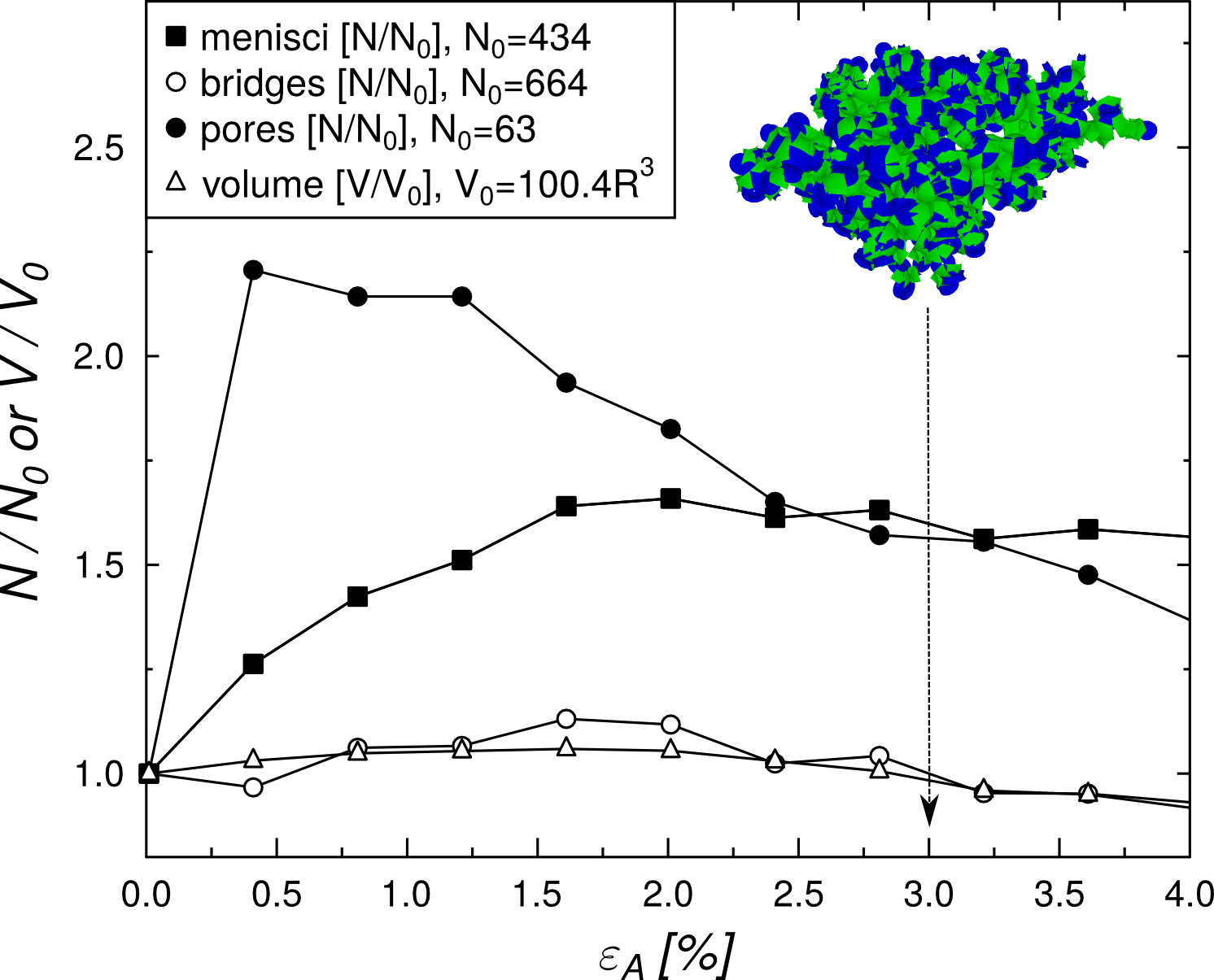}
\caption{Development of characteristic properties describing a single cluster as 
a function of the axial strain $\varepsilon_A$: volume of the cluster, menisci, 
bridges and filled pore bodies. Inset: liquid body of the studied cluster at 
$\varepsilon_A=3\%$. Bridges are colored blue and menisci are green. 
\label{Fig:singleClusterStatistics}}
\end{figure}

The deformation does not only have an effect on the statistics of liquid 
clusters. In particular for large deformations, parameters that characterize the 
cluster morphology evolve. We exemplify the morphogenesis of a single liquid 
cluster during the triaxial shear test ($W_c=22\%;~\sigma_3=40 u_p$).  While it 
remains rather isochoric with $V \approx 100 R^3$, there are changes in the 
number of individual units which constitute the cluster as the axial strain 
$\varepsilon_A$ increases (see  Fig.~\ref{Fig:singleClusterStatistics}). This is 
observed for the number of individual units like menisci, liquid bridges and 
filled pore bodies (see Sec.~\ref{sec:liquidbody}). The axial compression leads 
to spreading of the cluster in the plane vertical to the axial load, which can 
be seen from the increasing number of menisci. The increasing number of menisci 
can be explained by the formation of new trimer units due to local instabilities 
or the incorporation of already existing trimers by the large cluster after 
contacting. Note, that although we take a look at a single cluster here, also 
other (smaller) clusters exist in the sample. The number of filled pore bodies 
and the number of liquid bridges decrease after reaching a local maximum due to 
increasing inter-particle distances from dilatation in $\sigma_3$ directions 
that resulted in liquid bridge ruptures. The cluster evolves from a compact 
shape into a looser structure. The rapid increase of the number of menisci and 
filled pore bodies in the beginning (small strains $\epsilon_A$) does not result 
in a higher cluster volume because the Laplace pressure decreases when 
new pores are filled. A lower Laplace pressure leads to a 
decrease of menisci volumes since they move deeper into pore throats, see 
Fig.\ref{Fig:instab}.
\subsection{Dependence of Limit Surfaces on Liquid Content} \label{sec:mc_eval}
\begin{figure}[htb]
\centering
\includegraphics[width=0.5\textwidth]{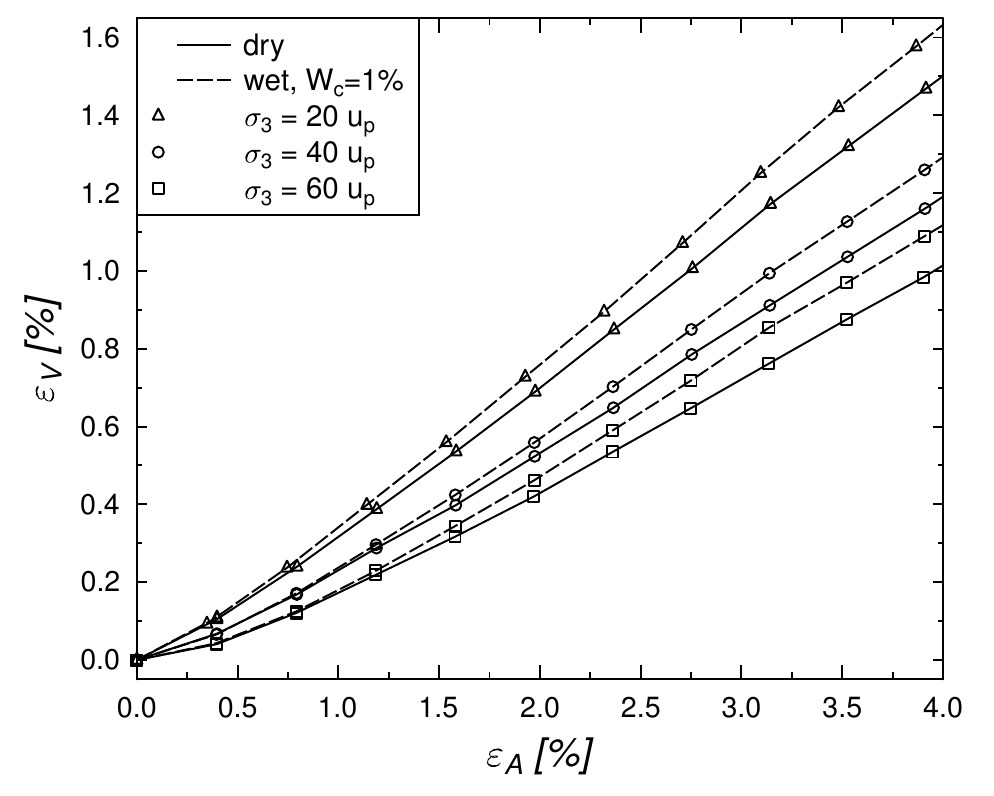}
\caption{Volumetric strain $\varepsilon_V$ as a function of axial strain 
$\varepsilon_A$ for different confining pressures $\sigma_3$. 
\label{Fig:volumetricStrain}}
\end{figure}

We previously referred to the role of the volumetric strain $\varepsilon_V$. Its 
dependence on the axial strain $\varepsilon_{A}$ and confining pressure 
$\sigma_3$ is shown in Fig.~\ref{Fig:volumetricStrain}. We observe a decrease of 
$\varepsilon_V$ with increasing $\sigma_3$, similar to 
Ref.~\cite{Scholtes2009a}. A typical observation is that the spread between 
volumetric strain curves for different confining pressures increases with 
increasing axial strain. Note that the spontaneous increase of 
$q=\sigma_1-\sigma_3$ is a consequence of the Contact Dynamics approach. 
Comparable studies with the soft particle contact models exhibit a slight 
initial decrease of the sample volume due to the elasticity of the particles 
(see e.g. Ref~\cite{Scholtes2009}). 
Attractive capillary forces hinder the spatial rearrangement of particles, 
resulting in a slightly higher volumetric strain for the wet sample
(see Fig.~\ref{Fig:volumetricStrain}).

\begin{figure}[htb]
\centering
\includegraphics[width=0.5\textwidth]{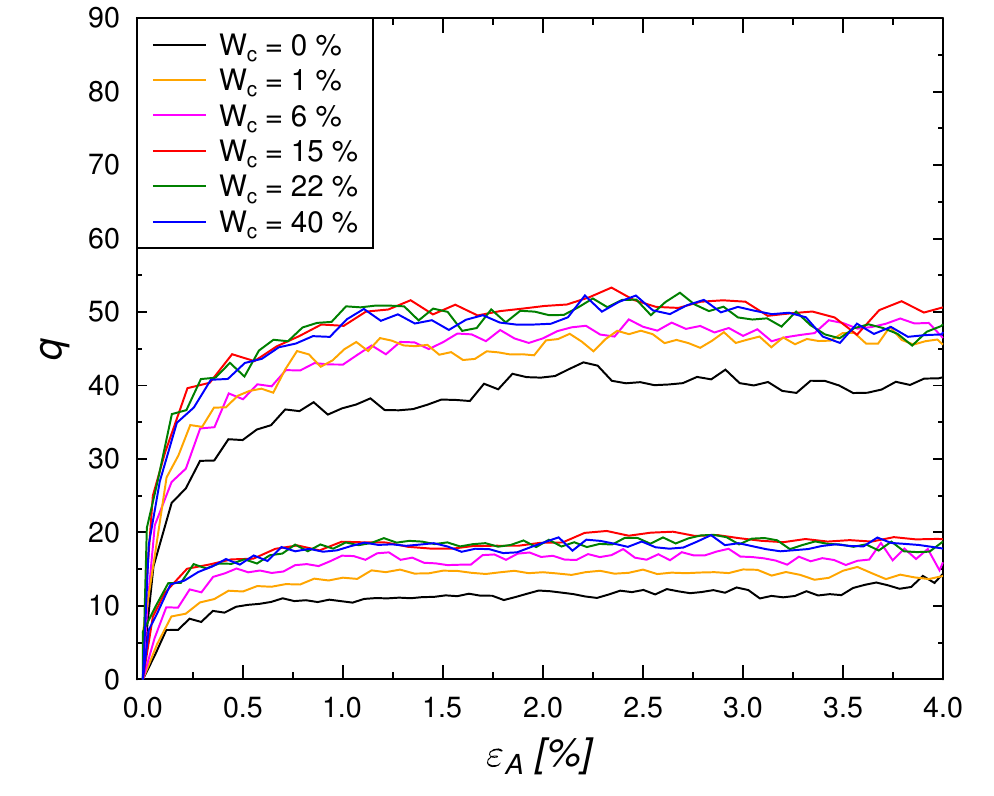}
\caption{Differential stress $q$ versus axial strain $\varepsilon_A$ for 
different liquid contents $W_c$. Two values of confining pressure $\sigma_3$ are 
considered: $\sigma_3=20 u_p$ (plateau in the range $7 u_p < q < 20 u_p$) and 
$\sigma_3=60 u_p$ (plateau in the range $35 u_p < q < 55 u_p$).}
\label{Fig:triaxialDiffSaturations}
\end{figure}

During the compression, the differential stress $q=\sigma_1-\sigma_3$ $[u_p]$ is 
calculated as function of the axial strain $\varepsilon_A$. In 
Fig.~\ref{Fig:triaxialDiffSaturations} the stress-strain curve exhibits a rapid 
increase for small $\varepsilon_A$ independent on the liquid content $W_c$. This 
behavior is typical for a dense packing and in agreement with experimental 
\cite{Wang2001} and numerical observations \cite{Belheine2008,Scholtes2009a}. 
For larger strains, $q$ saturates and reaches a constant plateau, whose value 
depends on $W_c$. 

\begin{figure}[htb]
\centering
\includegraphics[width=0.5\textwidth]{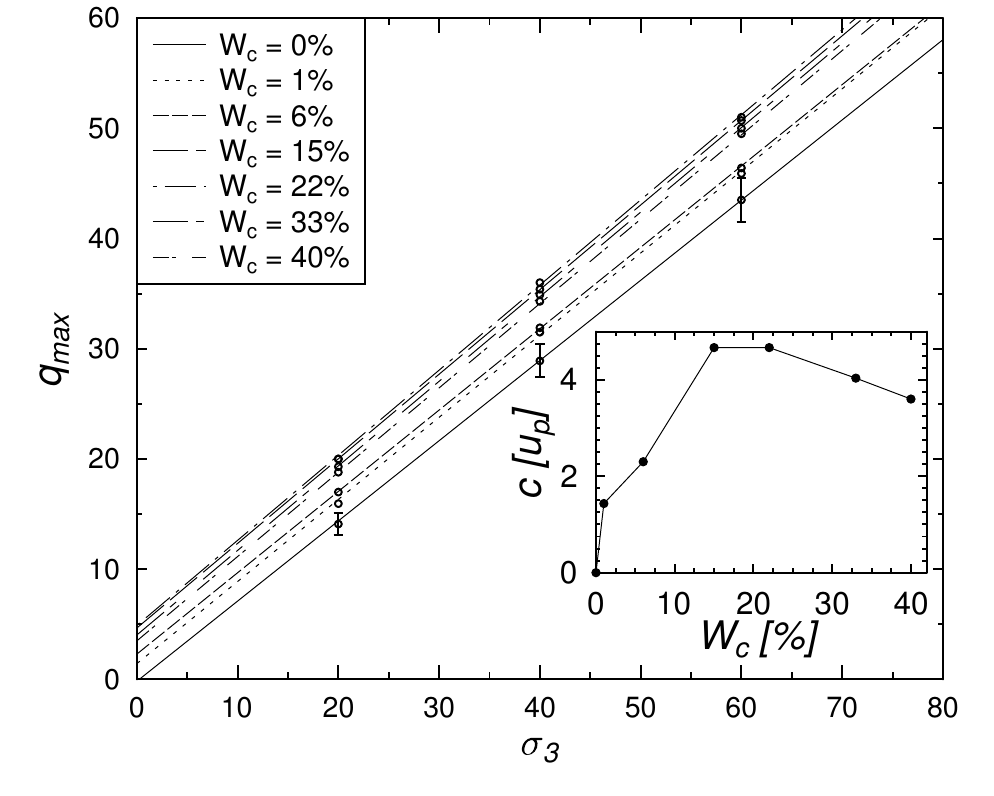}
\caption{Mohr-Coulomb failure envelope: maximal values of differential stress 
$q_{max}$ as a function of confining pressure $\sigma_3$. The data (single 
points) was extracted from the curves shown in Fig. 
\ref{Fig:triaxialDiffSaturations} by taking an average between 
$1.5\%<\epsilon_A<3.5\%$. Straight lines are the fit curves to the single data 
points.  Error bars for $W_c=0\%$ exemplary visualize the uncertainty due to 
fluctuations of $q_{max}$ in the numerical simulation which increase with 
$\sigma_3$. Inset: Cohesion $c$ as a function of liquid content $W_c$, extracted 
from the intersection of the lines with the $q_{max}$ axis at $\sigma_3=0$. 
\label{Fig:MohrCoulomb}}
\end{figure}

When monitoring the shear strength, or maximum in differential stress $q_{max}$ 
as function of the liquid content $W_c$ (see 
Fig.~\ref{Fig:triaxialDiffSaturations}) at a single value of confining pressure, 
one can see a huge increase from the dry state to $W_c=1\%$, corresponding to a 
pendular state where only liquid bridges exist. From there on, the gradient in 
shear strength with respect to liquid content quickly decays. Our data suggests 
a maximum in shear strength between $15\%\leq W_c \leq 22\%$, followed by a 
decay that in principle should reach the one of the dry state for $W_c=100\%$. 
Shear strength versus confining pressure can be approximated by a straight line 
which goes through zero for the dry granular material (see 
Fig.~\ref{Fig:MohrCoulomb}). This behavior is well known \cite{Scholtes2009} and 
described by the Mohr-Coulomb failure criterion in the form
\begin{equation}
 \tau = \mu_{f} \sigma + c,
\end{equation}
where $\tau$ is the shear strength, $\mu_{f} = \tan(\psi)$ the internal friction 
coefficient, $\psi$ denotes the angle of internal friction and $c$ cohesion. The 
curves for wet material also show a linear trend, but with cohesion that results 
from the sum of capillary forces of all liquid bridges and clusters. When 
looking at the dependence of $c$ on the liquid content $W_c$ (see 
Fig.~\ref{Fig:MohrCoulomb}(inset)) it becomes evident that the increased shear 
strength is due to an increased cohesion and not to a change in $\psi$. As a 
matter of fact $\psi$ proves to be independent on the liquid content within our 
statistical error bars at $\psi=35.9^{\circ}$.  The maximal increase in shear 
strength is about 5$u_p$, independent on the confining pressures.
\section{Summary and Conclusions}\label{sec:Conclusions}
The described micro-mechanical model couples liquid structures to individual 
moving particles. For simplicity, the solid phase consists of spherical 
particles whose evolution in time and space is simulated with Contact Dynamics. 
For the liquid phase, the model incorporates rather complicated liquid 
structures observed in experiments which became feasible due to recent advances 
in the micro-tomography technique \cite{Scheel2008nature}. To ensure the 
conservation of mass, volume is chosen as the control variable in the model. The 
corresponding pressure is calculated separately for each liquid structure. The 
fluid-particle coupling induces cohesive forces exerted by the liquid structures 
on the holding particles. Note, that the present version of the model is limited 
to small strains up to 4-5$\%$ where an initial triangulation structure does not 
need to be updated. For higher strains the particle positions must be 
triangulated anew to provide a valid pore network. The proposed model was 
applied for the simulation of triaxial compression tests at different saturation 
levels. We could qualitatively reproduce results reported in both numerical and 
experimental studies, including the well-known Mohr-Coulomb failure criterion. 
Furthermore, shear tests well beyond the pendular regime were simulated. Due to 
the micro-mechanical approach, we were able to track and characterize the 
evolution of liquid structures inside the deforming granular material and to 
relate it to the deformation state. In the future we will extend the model to 
larger deformations.
\begin{acknowledgements}
The research leading to these results has received funding from the People 
Programme (Marie Curie Actions) of the European Union's Seventh Framework 
Programme FP7 under the MUMOLADE ITN project (Multiscale Modelling of Landslides 
and Debris Flow) with REA grant agreement n$^{\circ}$ 289911, as well as from 
the European Research Council Advanced Grant no.~319968-FlowCCS and the DFG 
under PiKo SPP 1486 HE 2732/11-3.
\end{acknowledgements}

\bibliographystyle{spbasic}      

\bibliography{./paper.bbl}
\end{document}